\documentclass[a4paper,11pt,dvips]{article}
\usepackage{graphicx}
\usepackage{amsmath}
\usepackage{amssymb}
\usepackage{latexsym}
\usepackage{color}
\usepackage{cite}
\usepackage{makeidx}
\usepackage{float}
\usepackage{multirow}
\usepackage[dvipdfm,colorlinks]{hyperref}
\newcommand{\bra}{\begin{array}}
\newcommand{\era}{\end{array}}
\newcommand{\beq}{\begin{equation}}
\newcommand{\eeq}{\end{equation}}
\newcommand{\bqr}{\begin{eqnarray}}
\newcommand{\eqr}{\end{eqnarray}}

\def\BC{\bb C}
\def\_\BC{\bbi C}

\def\( {\left(}
   \def\) {\right)}
\def\[ {\left[}
\def\] {\right]}
\def\no2 {{\textstyle{n\over 2}}}

\newcommand{\al}{\alpha}

\newcommand{\lb}{\label}

\begin{document}
\begin{titlepage}
\setcounter{page}{1}
\renewcommand{\thefootnote}{\fnsymbol{footnote}}

\begin{flushright}
\end{flushright}

\vspace{5mm}
\begin{center}

{\Large \bf {Tunneling of Massive Dirac Fermions in Graphene \\ through
Time-periodic Potential}}

\vspace{5mm}

 {\bf Ahmed Jellal\footnote{\sf ajellal@ictp.it --
a.jellal@ucd.ac.ma}}$^{a,b}$, {\bf Miloud Mekkaoui}$^{b}$, {\bf
El Bou\^{a}zzaoui Choubabi}$^{b,c}$
 and {\bf Hocine Bahlouli}$^{a,d}$

\vspace{5mm}

{$^a$\em Saudi Center for Theoretical Physics, Dhahran, Saudi Arabia}

{$^{b}$\em Theoretical Physics Group,  
Faculty of Sciences, Choua\"ib Doukkali University},\\
{\em PO Box 20, 24000 El Jadida, Morocco}

{$^{c}$\em Physics Department, Faculty Polydisciplinary, Sultan Moulay Slimane University,\\
23000 Beni Mellal, Morocco}

{$^d$\em Physics Department,  King Fahd University
of Petroleum $\&$ Minerals,\\
Dhahran 31261, Saudi Arabia}



\vspace{3cm}

\begin{abstract}
The energy spectrum of a graphene sheet subject to a single barrier potential having a time periodic oscillating height
and subject to a magnetic field is analyzed. The corresponding transmission is studied as function of the incident energy
and potential parameters. Quantum interference within the oscillating barrier has an important effect on quasiparticles tunneling.
In particular the time-periodic electrostatic potential generates additional sidebands at energies
$\epsilon + l\hbar \omega$ ($l=0,\pm 1, \cdots$) in the transmission probability originating from the photon absorption
or emission within the oscillating barrier. Due to numerical difficulties in truncating the resulting coupled channel
equations we limited ourselves to low quantum channels, i.e. $l=0,\pm 1$.

\vspace{3cm}



\end{abstract}
\end{center}
\end{titlepage}


\section{ Introduction}

{
Graphene \cite{Geim} is a single layer of carbon atoms arranged into a planar honeycomb
lattice. This system has attracted a considerable attention from both experimental and theoretical
researchers since its experimental realization in 2004 \cite{Novoselov1}. This is because of
its unique and outstanding mechanical,
electronic, optical, thermal and chemical properties \cite{Castro}. Most of these marvelous properties are due
to the apparently relativistic-like nature of its carriers, electrons behave as massless Dirac fermions in graphene systems.
In fact starting from the original tight-binding Hamiltonian describing  graphene it has been shown theoretically
that the low-energy excitations of graphene appear to be massless chiral Dirac fermions. Thus, in the continuum
limit one can analyze the crystal properties using the formalism of quantum electrodynamics in (2+1)-dimensions.
This similarity between condensed matter physics and quantum electrodynamics (QED) provides the opportunity
to probe many physical aspects proper to high energy physics phenomena in condensed matter systems. Thus, in this regard,
graphene can be considered as a test-bed laboratory for high energy relativistic quantum phenomena.

Quantum transport in periodically driven quantum systems is an important subject not only of academic value but also
for device and optical applications. In particular quantum interference within an oscillating time-periodic electromagnetic field
gives rise to additional sidebands at energies $\epsilon + l\hbar \omega$ $ (l=0,\pm1,\cdots)$ in the transmission probability originating from
the fact that electrons exchange energy quanta $\hbar \omega $ carried by photons of the oscillating field, $\omega $ being the frequency
of the oscillating field. The standard model in this context is that of a time-modulated scalar potential in a finite region of space.
It was studied earlier by Dayem and Martin \cite{Dayem} who provided the experimental evidence of photon assisted tunneling in experiments on superconducting
films under microwave fields. Later on  Tien and Gordon \cite{Tien} provided the first theoretical explanation of these experimental
observations. Further theoretical studies were performed later by many research groups, in particular Buttiker investigated the
barrier traversal time of particles interacting with a time-oscillating barrier \cite{Buttiker}.
{Wagner  \cite{Wagner-1} gave a detailed
treatment on photon-assisted tunneling through a strongly driven double
barrier tunneling diode and studied the transmission probability of
electrons traversing a quantum well subject to a harmonic driving force
\cite{Wagner-2} where transmission side-bands have been predicted.
Grossmann \cite{Grossmann}, on the other hand, investigated the
tunneling through a double-well perturbed by a monochromatic driving
force which gave rise to unexpected modifications in the tunneling
phenomenon.}

In \cite{ahsan} the authors studied  the chiral tunneling through a harmonically driven potential barrier in a graphene monolayer.
Because the charge carriers in their system are massless they described the tunneling effect as the Klein tunneling with high anisotropy.
For this, they determined the transmission probabilities for the central band and sidebands in terms of the incident angle of
the electron beam. Subsequently, they investigated the transmission probabilities for varying width, amplitude and frequency of the
oscillating barrier. They conclude that the perfect transmission for normal incidence, which has been reported for a static barrier,
persists for the oscillating barrier that is a manifestation of Klein tunneling in a time-harmonic potential.

The growing experimental interest in studying optical properties of electron transport in graphene subject to strong laser fields \cite{Jiang1}
motivated the recent upsurge in theoretical study of the effect of time dependent periodic electromagnetic field on electron spectra.
Recently it was shown that laser fields can affect the electron density of states and consequently the electron transport properties \cite{Calvo}.
{Electron transport in graphene generated by laser
irradiation was shown to result in subharmonic resonant enhancement
\cite{San-Jose}.
The analogy between spectra of Dirac fermions in laser fields and the
energy spectrum in graphene superlattice formed by static one
dimensional periodic potential was performed in \cite{Savelev-1}. In
graphene systems resonant enhancement of both electron backscattering
and currents across a scalar potential barrier of arbitrary space and
time dependence was investigated in \cite{Savelev-2} and resonant
sidebands in the transmission due to a time modulated potential region
was studied recently in graphene \cite{Liu}. The fact that an applied
oscillating field can
result in an effective mass or equivalently a dynamic gap was confirmed
in recent studies \cite{Fistul}. Adiabatic quantum pumping of a graphene
devise with two oscillating electric barriers was considered in
\cite{Evgeny}. A Josephson-like current was predicted for several time
dependent scalar potential barriers placed upon a monolayer of graphene
\cite{Savelev-3}. Stochastic resonance like phenomenon
\cite{Gammaitoni} was predicted for transport phenomena in disordered
graphene nanojunctions \cite{Jiang}. Further study showed that
noise-controlled effects can be induced due to the interplay between
stochastic and relativistic dynamics of charge carriers in graphene
\cite{Pototsky}.}

}

In this work we generalize the results obtained in \cite{ahsan} in the presence of a magnetic field case. More precisely, we consider
one monolayer graphene sheet lying in the $xy$-plane and subject to a scalar square potential barrier along the $x$-direction while
the carriers are free in the $y$-direction. The barrier height oscillates sinusoidally around an average value
$V_0$ with oscillation amplitude $V_1$ and frequency $\omega$. We
calculate the transmission probability for the central band and
close by sidebands as a function of the potential parameters and
incident angle of the particles. The limitation to close by
sidebands is due to numerical difficulties in truncating the
resulting coupled channel equations which forced us to limit
ourselves to low quantum channels.

The manuscript is organized as follows. In section 2, we present the theoretical
model describing the graphene sheet in the presence of an external magnetic field
and oscillating barrier potential. In section 3, we explicitly determine
 the eigenvalues and  corresponding eigenspinors for each regions
 composing ours system. We study the energy spectrum by investigating
 different properties to underline its behavior with respect to changes of
 physical parameters
 in section 4.  The transmission through oscillating barrier
 will be analyzed in section 5 followed by a
discussion of the numerical results in section 6. To complete our study,
we deal with the total transmission probability in section 7.
Our conclusions are given in the final section.

\section{ Theoretical model}

We study the tunneling effect of a system   of Dirac
fermions living in two-dimensions. This system is a flat sheet of graphene subject to a
square potential barrier along the $x$-direction while particles
are free in the $y$-direction. The width of the barrier is $d$,
its height is oscillating sinusoidally around $V_0$ with amplitude
$V_{1}$ and frequency $\omega$. The intermediate zone is subject
to a magnetic field $\textbf{B}=B(x, y)\textbf{e}_z$ perpendicular to the graphene sheet. Electrons with energy
$\epsilon={E}/{v_F}$ are incident from one side of the barrier with an angle
$\phi_{0}$ with respect to the $x$-direction and leaves the
barrier with energy $\epsilon+ l\hbar \omega$ $(l=0, \pm 1,
\cdots)$, which $l$ are the modes generated by oscillations and
making angles $\pi-\phi_{l}$ after reflection and $\theta_{l}$
after
transmission. The corresponding Hamiltonian 
can be split
into two parts
\begin{equation}\lb{ham1}
H=H_{I}+H_{II}
\end{equation}
such that the first one is
\begin{equation}\lb{ham2}
H_{I}=v_{F} {\boldsymbol{\sigma}} \cdot \left(-i\hbar
{\boldsymbol{\nabla}}+
\frac{e}{c}\textbf{A}(x,y)\right)+V(x){\mathbb I}_{2}
\end{equation}
and the second one describes the harmonic time dependence of
the barrier height
\begin{equation}
H_{II}=V_{j}\cos(\omega t)
\end{equation}
where $\upsilon_{F}$ is the Fermi velocity, $
{\boldsymbol{\sigma}} =(\sigma_{x}, \sigma_{y})$ are the  Pauli
matrices and ${\mathbb I}_{2}$ is the $2 \times 2$ unit matrix.
$V$ and $V_{j}$ are the static square potential barrier and the
amplitude of the oscillating potential, respectively. Both $V$ and
$V_{j}$ are constants for $0\leq x\leq d$ with $d$ positive and
are zero elsewhere, which can be summarized as
\begin{equation}
V(x)=
\left\{%
\begin{array}{ll}
    V_{0}, & \qquad\hbox{$0\leq x\leq d$} \\
    0, & \qquad \hbox{otherwise} \\
\end{array}%
\right.,\qquad V_{j}=
\left\{%
\begin{array}{ll}
    V_{1}, & \qquad \hbox{$0\leq x\leq d$} \\
    0, & \qquad \hbox{otherwise} \\
\end{array}%
\right.
\end{equation}
and the script $j = {\sf 0}, {\sf 1}, {\sf 2}$ denotes each
scattering region. For a magnetic barrier, the relevant physics is
described by a magnetic field translationally invariant along the
$y$-direction, $B(x, y)= B(x)$. Choosing the Landau gauge we
impose the vector potential  ${\boldsymbol{A}} =
(0,A_{y}(x))^{T}$ with $\partial_{x}A_{y}(x)= B(x)$, the
transverse momentum $p_{y}$ is thus conserved. The magnetic field
$\textbf{B}=B_0\textbf{e}_z$ (with constant $B_{0}$) within the
strip $0\leq x\leq d$ but $B=0$ elsewhere, such as
\begin{equation}
B(x,y)= B_{0}\Theta(dx-x^{2})
\end{equation}
with the Heaviside step function $\Theta$
\begin{equation}
\Theta(x)=\left\{%
\begin{array}{ll}
    1, & \qquad \hbox{$x>0$} \\
    0, & \qquad \hbox{otherwise.} \\
\end{array}%
\right.
\end{equation}
\begin{figure}[H]
\centering
\includegraphics[width=11
cm,height=6cm]{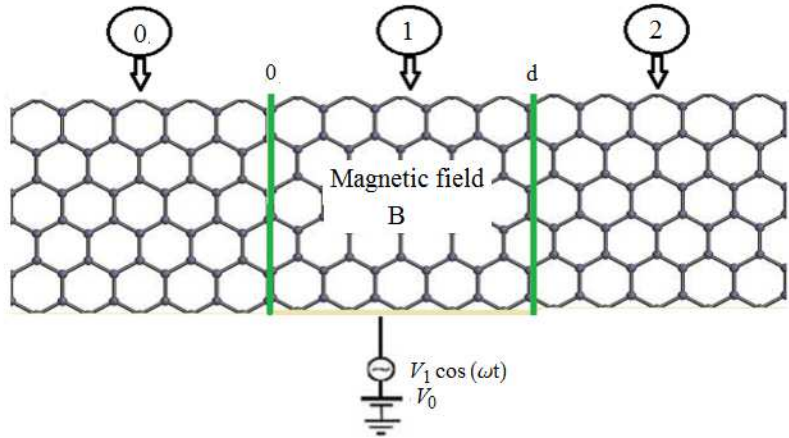}\\
 \caption{\sf{
  { (Color online) Schematic of  an oscillating potential in a magnetic field of the monolayer graphene}.}}
\end{figure}
\noindent The continuity of the corresponding potential vector takes the
following expression
\begin{equation}
\qquad A_{y}(x)=\left\{%
\begin{array}{ll}
    0, & \qquad \hbox{$x<0$} \\
    B_{0}x, & \qquad \hbox{$0\leq x\leq d$} \\
    B_{0}d, & \qquad \hbox{$x>d$}. \\
\end{array}%
\right.
\end{equation}
Our system can be presented in Figure 1
to show clearly the oscillating potential in a magnetic field of
the monolayer. On the light of this, we
present in Figure 2 how the electrons can be scattered by our barrier
potential. This will help to analyze the tunneling effect and
calculate different physical quantities.
\begin{figure}[H]
\centering
\includegraphics[width=10
cm,height=7cm]{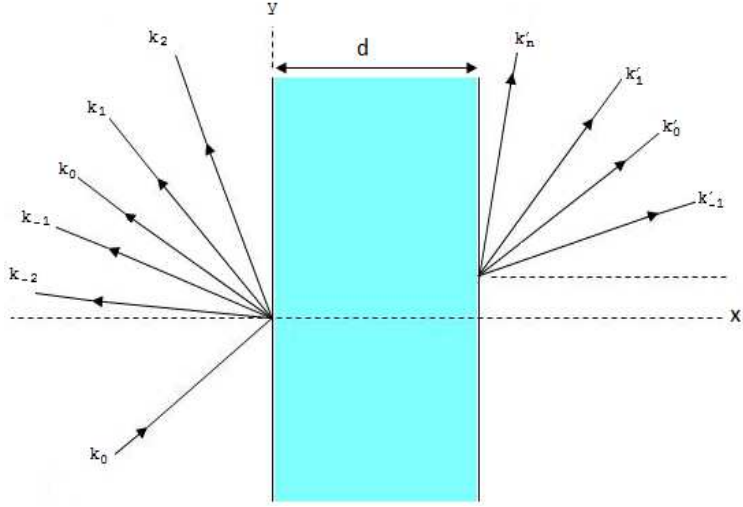}\\
 \caption{\sf{(Color online)
  Geometry of electron diffraction.}}
\end{figure}

\section{Energy spectrum }

To explicitly determine the solutions of the energy spectrum of our
theoretical model, we separately handle each part of the Hamiltonian \eqref{ham1}.
Thus, let us start from the
time-dependent Dirac equation in the absence of oscillating
potential for the spinor $\psi(x, y)=(\psi_{+},\psi_{-})^{T}$ at
energy $E$. This is
\begin{equation}
H_{I}\psi(x,y,t)= E \psi(x,y,t)
\end{equation}
where
$\psi(x , y, t)= \psi(x, y)e^{-iEt/\hbar}$.
In matrix form,  we have 
\begin{equation}\lb{eq9}
\left(%
\begin{array}{cc}
  0 & -i\partial_{x}-\partial_{y}-\frac{ie}{\hbar c}A(x) \\
  -i\partial_{x}+\partial_{y}+\frac{ie}{\hbar c}A(x) & 0 \\
\end{array}%
\right)\left(%
\begin{array}{c}
  \psi_{+} \\
  \psi_{-} \\
\end{array}%
\right)=\frac{E}{\hbar \upsilon_{F}}\left(%
\begin{array}{c}
  \psi_{+} \\
  \psi_{-} \\
\end{array}%
\right).
\end{equation}
Since the transverse momentum $p_{y}$ is conserved, we can write
the wave function in separable form {$\psi_{\pm}(x,
y)=\varphi_{\pm}(x)e^{ik_{y}y}$}. Thus after rescaling energy
$\epsilon = E/v_{F}$ and using the unit system with $(\hbar= c = e
= 1)$, we obtain the two linear differential equations
\begin{eqnarray}
  && \left (-i\partial_{x}-ik_{y}-i A(x) \right)\varphi_{-}=\epsilon \varphi_{+}\\
&&   \left (-i\partial_{x}+ik_{y}+i A(x)
\right)\varphi_{+}=\epsilon \varphi_{-}.
\end{eqnarray}
This can be combined to describe the solution of \eqref{eq9} and
then  consider the incoming electrons to be in plane wave states
$\psi_{inc}(x,y,t)$ at energy $\epsilon$ as
\begin{equation}
\psi_{inc}(x,y,t)=\left(
\begin{array}{c}
1 \\
 \alpha_{0}\end{array}\right)e^{ik_{0}x}e^{ik_{y}y}e^{-iv_{F}\epsilon t}
\end{equation}
where $\alpha_{0}$ is given by
\begin{equation}
\alpha_{0}=s_{0}\frac{k_{0} +ik_{y}}{\sqrt{k_{0}^{2}
+k_{y}^{2}}}=s_{0} e^{\textbf{\emph{i}}\phi_{0}}
\end{equation}
with $s_{0}=\mbox{sgn}(\epsilon)$, $\phi_{0}$ is the angle that
the incident electrons make with the {$x$-direction},
$k_{0}$ and $k_{y}$ are the $x$ and $y$-components of the electron
wave vector, respectively. After rescaling the potential $v_{j}=
V_{j}/v_{F}$ and frequency $\varpi=\omega/v_{F}$, we show that the
transmitted and reflected waves have components at all energies
$\epsilon+ l\varpi$ $(l=0, \pm 1, \cdots)$. Indeed the wave
functions $\psi_{r}(x,y,t)$ for reflected electrons are
\begin{equation}
\psi_{r}(x,y,t)=\sum^{+\infty}_{m,l=-\infty} r_{l}\left(
\begin{array}{c}
1 \\
 -\frac{1}{\alpha_{l}}\end{array}\right)e^{-ik_{l} x +ik_{y}
 y} J_{m-l} \left(\frac{v_{j}}{\varpi}\right)\ e^{-iv_{F}(\epsilon+m\varpi)t}
\end{equation}
and the corresponding energy reads as
\begin{equation}\lb{energy1}
\epsilon+l\varpi=s_{l}\sqrt{k^{2}_{l}+k^{2}_{y}}
\end{equation}
where $r_{l}$ is the reflection amplitude and
$J_{m}\left(\frac{v_{1}}{\varpi}\right)$ is the Bessel function of
the first kind. Note that,
for the modulation amplitude  $v_{j} = 0$ we have
 $J_{m-l}
\left(0\right)=\delta_{ml}$. We will return to this point once we
talk about the solution in different regions composing the
graphene sheet. The parameter $\al_l$ is the complex number
\begin{equation}
\alpha_{l}=s_{l}\frac{k_{l} +ik_{y}}{\sqrt{k^{2}_{l}
+k_{y}^{2}}}=s_{l}\  e^{\textbf{\emph{i}}\phi_{l}}
\end{equation}
where $\phi_{l}=\tan^{-1}(k_{y}/k_{l})$,
$s_{l}=\mbox{sgn}(\epsilon+l \varpi)$, the sign again refers to
conduction and valence bands of region. The
{wave vector $k_l$ for mode $l$} can be obtained
from \eqref{energy1}
\begin{equation}
k_{l}=s_{l}\sqrt{\left(\epsilon+l\varpi\right)^{2}-k^{2}_{y}}.
\end{equation}
While, the wave functions $\psi_{t}(x,y,t)$ for transmitted
electrons read as
\begin{equation}
\psi_{t}(x,y,t)=\sum^{+\infty}_{m,l=-\infty}t_{l}\left(
\begin{array}{c}
1 \\
 \beta_{l}\end{array}\right)e^{ik^{'}_{l} x +ik_{y}
 y}J_{m-l} \left(\frac{v_{j}}{\varpi}\right) e^{-iv_{F}(\epsilon+m\varpi)t}
\end{equation}
and the eigenvalues are
\begin{equation}
\epsilon+l\varpi=s_{l}
\sqrt{k^{'2}_{l}+\left(k_{y}+\frac{d}{l^{2}_{B}}\right)^{2}}
\end{equation}
where $l_{B}=\sqrt{1/B_{0}}$ is the magnetic length , $t_{l}$ is the transmission
amplitude  and different parameters are given by
\begin{eqnarray}
&&
\beta_{l}=s_{l}\frac{k^{'}_{l}
+i\left(k_{y}+\frac{d}{l^{2}_{B}}\right)}{\sqrt{k^{'2}_{l}
+\left(k_{y}+\frac{d}{l^{2}_{B}}\right)^{2}}}=s_{l}\
e^{\textbf{\emph{i}}\theta_{l}}\\
&&
k^{'}_{l}=s_{l}\sqrt{\left(\epsilon+l\varpi\right)^{2}-\left(k_{y}+\frac{d}{l^{2}_{B}}\right)^{2}}\\
&&
\theta_{l}=\tan^{-1}\left[\left(k_{y}+\frac{d}{l^{2}_{B}}\right)/k^{'}_{l}\right].
\end{eqnarray}
At this level we summarize our solutions by writing the scattering states in
different regions. Recall that, in regions  ${\sf 0}$ and  ${\sf 2}$ the potential
height is $v_j=0$, then we proceed by replacing $J_{m-l}$ by $\delta_{ml}$. Consequently,
in region ${\sf 0}$, i.e. $x<0$, we have
\begin{equation}
\psi_{\sf
0}(x,y,t)=e^{ik_{y}y}\sum^{+\infty}_{m,l=-\infty}\left[\delta_{l0} \left(
\begin{array}{c}
1 \\
 \alpha_{l}\end{array}\right)e^{ik_{l}x}+r_{l}\left(
\begin{array}{c}
1 \\
 -\frac{1}{\alpha_{l}}\end{array}\right)e^{-ik_{l} x
 }\right]\delta_{ml}\  e^{-iv_{F}(\epsilon+m\varpi)t}
\end{equation}
and region ${\sf 2}$ $(x>d)$
\begin{equation}
\psi_{\sf
2}(x,y,t)=e^{ik_{y}y}\sum^{+\infty}_{m,l=-\infty}\left[t_{l}\left(
\begin{array}{c}
1 \\
 \beta_{l}\end{array}\right)e^{ik^{'}_{l} x
 }+b_{l}\left(
\begin{array}{c}
1 \\
 -\frac{1}{\beta_{l}}\end{array}\right)e^{-ik^{'}_{l} x}\right]\delta_{ml}\
  e^{-iv_{F}(\epsilon+m\varpi)t}
\end{equation}
where $\{b_{l}\}$ is a set of the null vectors.

In the barrier region ${\sf
1}$ $(0 \leq x \leq d)$, where $H_{II}$ is non-zero, the
eigenfunctions $\psi_{\sf 1}(x,y,t)$ of the total Hamiltonian $H$ can be expressed in
terms of the eigenfunctions $\psi_{1}(x,y)$ at energy $\epsilon$
of $H_{I}$. These are given by
\begin{equation}\lb{state1}
{\psi_{\sf 1}(x,y,t)=\psi_{1}(x,y)\sum^{+\infty}_{m=-\infty} J_{m}
\left(\alpha\right)\ e^{-iv_{F}(\epsilon+\varpi m)t}}
\end{equation}
{where we have set $\alpha={v_{1}}/{\varpi}$}. To include all
modes, a linear combination of wave functions at energies
$\epsilon_{l}=\epsilon+l\varpi$ $(l=0,\pm 1,\cdots)$ has to be
taken. Hence, one has to write \eqref{state1} as
\begin{equation}
 \psi_{\sf 1}(x,y,t)=\sum^{+\infty}_{l=-\infty}\psi_{l}(x,y)
 \sum^{+\infty}_{m=-\infty} J_{m} \left(\alpha\right)\ e^{-iv_{F}(\epsilon+\varpi(l+m))t}
\end{equation}
where eigenspinors
$\psi_{l}(x,y)$ are solution of the following equation
\begin{equation}
\left[{\boldsymbol{\sigma}}\cdot
{\boldsymbol{\pi}_{l}}+\frac{1}{v_{F}}V {\mathbb
I}_{2}\right]\psi_{l}(x,y)=\epsilon_{l}\psi_{l}(x,y)
\end{equation}
 with $\pi_{lx}=p_{lx}$ and $\pi_{y}=p_{y}+A_{y}$. The
$y$-component of the momentum is a constant of motion and the
spinor wave function can be written as
 $\psi_{l}(x,
y)=\varphi_{l}(x)e^{ik_{y}y}.$
We solve the eigenvalue equation for a given spinor
$\varphi_{l}=(\varphi_{l,1}, \varphi_{l,2})^{T}$
\begin{equation}\lb{mstate}
\left(%
\begin{array}{cc}
  V/v_{F} & -i
  \left(\partial_{lx}+k_{y}+\frac{x}{l_{B}^{2}}\right)\\
 i\left(-\partial_{lx}+k_{y}
 +\frac{x}{l_{B}^{2}}\right)  &  V/v_{F}\\
\end{array}%
\right)\left(%
\begin{array}{c}
  \varphi_{l,1} \\
  \varphi_{l,2}\\
\end{array}%
\right)=(\epsilon+l\varpi)\left(%
\begin{array}{c}
  \varphi_{l,1} \\
  \varphi_{l,2} \\
\end{array}%
\right).
\end{equation}
Defining the usual bosonic operators
\begin{eqnarray}
a_{l}=\frac{l_{B}}{\sqrt{2}}
\left(\partial_{lx}+k_{y}+\frac{x}{l_{B}^{2}} \right), \qquad
 a_{l}^{\dagger}=\frac{l_{B}}{\sqrt{2}}
\left(-\partial_{lx}+k_{y}+\frac{x}{l_{B}^{2}} \right)
\end{eqnarray}
which satisfy the commutation relation $[a_{l},
a_{k}^{\dagger}]=\delta_{lk}$. Rescaling our potential
$v=V_{0}/v_{F}$, in terms of $a_{l}$ and $a_{l}^{\dagger}$
\eqref{mstate} reads as
\begin{equation}
 \left(%
\begin{array}{cc}
  v& -i\frac{\sqrt{2}}{l_{B}}a_{l} \\
  i\frac{\sqrt{2}}{l_{B}}a_{l}^{\dagger}  &  v \\
\end{array}%
\right)\left(%
\begin{array}{c}
  \varphi_{l,1} \\
  \varphi_{l,2} \\
\end{array}%
\right)=(\epsilon+l\varpi)\left(%
\begin{array}{c}
  \varphi_{l,1} \\
  \varphi_{l,2}\\
\end{array}%
\right)
\end{equation}
which gives two relations between
spinor components
\begin{eqnarray}
 && -i\frac{\sqrt{2}}{l_{B}} a_{l}\varphi_{l,2}=(\epsilon+l\varpi-v)\varphi_{l,1} \lb{feq}\\
  && i\frac{\sqrt{2}}{l_{B}}a_{l}^{\dagger}\varphi_{l,1}=(\epsilon+l\varpi-v)\varphi_{l,2} \lb{seq}.
\end{eqnarray}
Now injecting \eqref{seq} in \eqref{feq}, we obtain a differential
equation of second order for $\varphi_{l,1}$
\begin{equation}
(\epsilon+l\varpi-v)^{2}\varphi_{l,1}=\frac{2}{l_{B}^{2}} a_{l}
a_{l}^{\dagger}\varphi_{l,1}.
\end{equation}
It is clear that $\varphi_{l,1}$ is an eigenstate of the number
operator ${N_{l}}=a_{l}^{\dagger}a_{l}$ and therefore we 
identify $\varphi_{l,1}$ with the eigenstates of the harmonic
oscillator 
\begin{equation}
 \varphi_{l,1} \sim \mid n_{l}-1\rangle
\end{equation}
and the corresponding eigenvalues are given by 
\begin{equation}\lb{epsl}
\epsilon_{l}=\epsilon+l\varpi=v\pm\frac{1}{l_{B}}\sqrt{2n_{l}}.
\end{equation}
 The second spinor component can be obtained
from \eqref{seq} to end up with
\begin{equation}
\varphi_{l,2}=
       \frac{i\sqrt{2n_{l}}}{\epsilon l_{B}+l\varpi l_{B}-vl_{B}} \mid n_{l}\rangle
\varphi_{l,2}= \pm i \mid n_{l}\rangle
\end{equation}
Thus, combining all to get 
the eigenspinors
\begin{equation}
\varphi_{l}^{\pm}=\left(%
\begin{array}{c}
   \mid n_{l}-1\rangle \\
  \pm i \mid n_{l}\rangle \\
\end{array}%
\right)
\end{equation}
where the wave functions $ \varphi_{n_{l}}(x)=\langle x\mid n_{l}\rangle$
can be written in terms of
the parabolic cylinder
functions $D_{n_{l}}(Q)$ as 
%
 \begin{equation}
\varphi_{n_{l}}(x)=c_{n_{l}}D_{n_{l}}(Q), \qquad D_{n_{l}}(Q)=2^{-\frac{n_{l}}{2}}e^{-\frac{Q^{2}}{4}}
H_{n_{l}}\left(\frac{Q}{\sqrt{2}}\right)
\end{equation}
 and 
and $H_{n_{l}}$ are the Hermite functions,
$Q=\sqrt{2}\frac{x+x_{0}}{l_{B}}$ and
$c_{n_{l}}=1/\sqrt{n_{l}!l_{B}^{2}\pi}$
which satisfy the recurrence
relation
$c_{n_{l}}=\frac{1}{\sqrt{n_{l}}}c_{n_{l}-1}.$
Finally, the solution in region ${\sf II}$ can be expressed
as 
\begin{eqnarray}
\psi_{\sf 1}(x,y,t) = e^{ik_{y}y}\sum^{+\infty}_{m,l=-\infty}\sum_{\pm}c^{\pm}_{l}\left(%
\begin{array}{c}
 D_{1/\left(\Lambda_{l}\right)^{2}-1}
 \left[\pm \sqrt{2}\left(\frac{x}{l_{B}}+k_{y}l_{B}\right)\right] \\
  \pm i\Lambda_{l}
  D_{1/\left(\Lambda_{l}\right)^{2}}
  \left[\pm \sqrt{2}\left(\frac{x}{l_{B}}+k_{y}l_{B}\right)\right] \\
\end{array}%
\right)
J_{m-l}\left(\alpha\right)\ e^{-iv_{F}(\epsilon+m\varpi)t }
\end{eqnarray}
where we have set
\begin{equation}
\Lambda_{l}=\frac{\sqrt{2}}{ l_{B}|\epsilon +l\varpi-v|}.
\end{equation}

Having obtained all solutions of the energy spectrum, we will see how they can
be used to deal with different issues. Specifically, the determination
of transmission and reflection in terms of the different physical parameters
of our system.

\section{Spectrum properties}

At this level let us study our eigenvalues to
underline their basic features. From \eqref{epsl}, we obtain
the energy modulation due the oscillating potential
as shown in Figure 3:

\begin{figure}[H]
\centering
\includegraphics[width=10
cm,height=4cm]{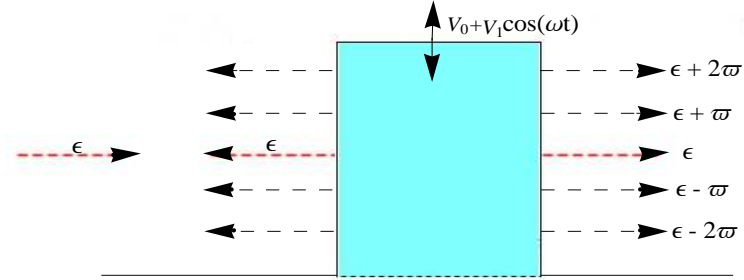}\\
 \caption{\sf{ (Color online)
  Schematic oscillating barrier.}}
\end{figure}
\noindent Figure 3 gives an idea how the energy spectrum looks
like. It is clearly seen that absorbing energy quantum �$\varpi$ produces
interlevel transitions. Because of the Pauli principle an electron
with energy $\epsilon$ can absorb an energy quantum �$\varpi$ if
only the state with energy $\epsilon+\varpi$ is empty. 
After absorbing the state with energy $\epsilon$
becomes empty, then one can write
\begin{equation}
\epsilon=v-l\varpi\pm\frac{1}{l_{B}}\sqrt{2n_{l}}
\end{equation}
and from \eqref{epsl} we can each time fix $l$ to end up
with the set of energies 
\begin{equation}
\epsilon_{0}=\epsilon=v\pm\frac{1}{l_{B}}\sqrt{2n_{0}},
\qquad
\epsilon_{1}=\epsilon+\varpi=v\pm\frac{1}{l_{B}}\sqrt{2n_{1}}, \qquad
\epsilon_{2}=\epsilon+2\varpi=v\pm\frac{1}{l_{B}}\sqrt{2n_{2}}\  \cdots.
\end{equation}
Note that, the energy conservation imposes the condition
\begin{equation}
\epsilon=v\pm\frac{1}{l_{B}}\sqrt{2n_{0}}=-\varpi+v\pm\frac{1}{l_{B}}\sqrt{2n_{1}}
=-2\varpi+v\pm\frac{1}{l_{B}}\sqrt{2n_{2}}= \cdots =-l\varpi+v\pm\frac{1}{l_{B}}\sqrt{2n_{l}}
\end{equation}
which implies that the energy for any integer value $l$ can be written as
\beq
\epsilon_{l}=v+l\varpi\pm\frac{1}{l_{B}}\sqrt{2n_{0}}.
\eeq
It is clearly seen that the difference of energy is
$\epsilon_{l+1}-\epsilon_{l}=\varpi$, which independent of the quantum numebr $n_0$.
Combining all to present the energy in terms of the external magnetic field in Figure 4. One can notice that
for $n_0=0$, we have just modulation of the energy with different number quanta $l\varpi$ with $l=0,\pm1, \cdots$. However for $n_0=1,2$ the energy
behavior is completely changed and for each $l$ value the energy is split into two values, which can be seen like
a left of degeneracy of levels.
\begin{figure}[H]
\centering
\includegraphics[width=10
cm,height=7cm]{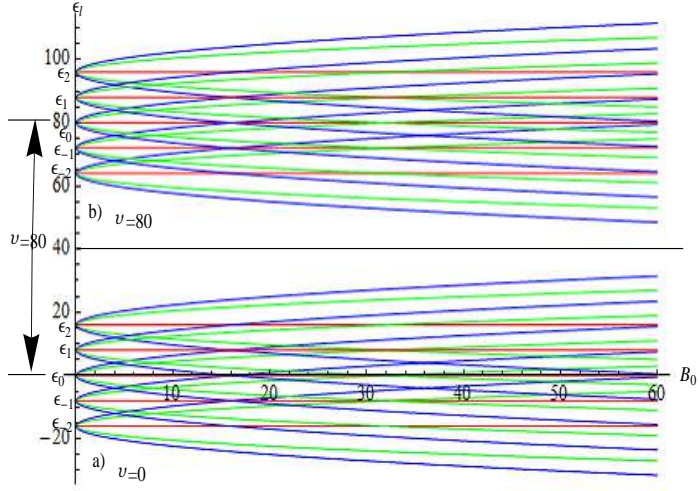}\\
 \caption{\sf{{(Color online)}
  Graphs depicting the energy  $\epsilon_{l}$ as a function of magnetic field
  $B_{0}$,  
  $\varpi=8$, potential $v=\{0,80\}$,
    $n_{0}=0$ (red),
$n_{0}=1$ (green)  and $n_{0}=2$ (blue).}}
\end{figure}

\section{Transmission through oscillating barrier}

Note that for our system, as Dirac electrons pass through a region subjected to
time-harmonic potentials, transitions from the central band to
sidebands (channels) at energies $\epsilon+l\varpi$ $(l = 0, \pm 1, \cdots)$
occur as electrons ex-change energy quanta with the
oscillating field. Then to handle the propagation of waves, we need
 the transmission and reflection amplitudes,
which can be determined by matching different wave functions
at interfaces $0$ and $d$ to write
\begin{eqnarray}
  &&  \psi_{\sf 0}(0,y,t)=\psi_{\sf 1}(0,y,t) \lb{psix0}\\
   && \psi_{\sf 1}(d,y,t)=\psi_{\sf 2}(d,y,t) \lb{psixd}.
\end{eqnarray}
For simplify of writing, we use the shorthand notation
\begin{eqnarray}
&&\eta_{1,l}^{\pm}=D_{1/\left(\Lambda_{l}\right)^{2}-1}
 \left(\pm \sqrt{2}k_{y}l_{B}\right)\\
&& \xi_{1,l}^{\pm}= D_{1/\left(\Lambda_{l}\right)^{2}}
  \left(\pm \sqrt{2}k_{y}l_{B}\right)\\
&& \eta_{2,l}^{\pm}=D_{1/\left(\Lambda_{l}\right)^{2}-1}
 \left[\pm \sqrt{2}\left(\frac{d}{l_{B}}+k_{y}l_{B}\right)\right]\\
&& \xi_{2,l}^{\pm}= D_{1/\left(\Lambda_{l}\right)^{2}}
  \left[\pm \sqrt{2}\left(\frac{d}{l_{B}}+k_{y}l_{B}\right)\right].
\end{eqnarray}
To derive different physical quantities, one can explicitly write (\ref{psix0}-\ref{psixd})
by making use
 the fact
 that the basis $\{e^{imv_{F}\varpi t}\}$ is orthogonal. Thus
at interface $x=0$, one finds
\begin{eqnarray}
&& \delta_{m0}+r_{m}=\sum^{+\infty}_{l=-\infty}
\left(c^{+}_{l}\eta_{1,l}^{+} +c^{-}_{l}\eta_{1,l}^{-}\right)
J_{m-l}\left(\alpha\right) \lb{eqx01}\\
&& \delta_{m0}\alpha_{m}-r_{m}\frac{1}{\alpha_{m}}=
\sum^{+\infty}_{l=-\infty}
\left(c^{+}_{l}i\Lambda_{l}\xi_{1,l}^{+}-c^{-}_{l}i\Lambda_{l}\xi_{1,l}^{-}\right)
J_{m-l}\left(\alpha\right) \lb{eqx02}
\end{eqnarray}
 at $x=d$ we have 
\begin{eqnarray}
&& t_{m}e^{ik^{'}_{m}d}+
b_{m}e^{-ik^{'}_{m}d}=\sum^{+\infty}_{l=-\infty} \left(
 c^{+}_{l}\eta_{2,l}^{+} +c^{-}_{l}\eta_{2,l}^{-}\right) J_{m-l}\left(\alpha\right) \lb{eqxd1} \\
&&t_{m}\beta_{m}e^{ik^{'}_{m}d}-b_{m}\frac{1}{\beta_{m}}e^{-ik^{'}_{m}d}
= \sum^{+\infty}_{l=-\infty}
\left(c^{+}_{l}i\Lambda_{l}\xi_{2,l}^{+} -c^{-}_{l}i\Lambda_{l}
\xi_{2,l}^{-}\right)J_{m-l}\left(\alpha\right). \lb{eqxd2}
\end{eqnarray}

 It is convenient to write (\ref{eqx01}-\ref{eqxd2}) in matrix form, such as 
\begin{eqnarray}
\left(%
\begin{array}{c}
  \Xi_{0} \\
  \Xi_{0}^{'} \\
\end{array}%
\right)=\left(%
\begin{array}{cc}
 { \mathbb M11} &{\mathbb M12} \\
 {\mathbb M21} &{ \mathbb M22} \\
\end{array}%
\right)\left(%
\begin{array}{c}
  \Xi_{2} \\
  \Xi_{2}^{'}\\
\end{array}%
\right)={\mathbb M}\left(%
\begin{array}{c}
  \Xi_{2} \\
 \Xi_{2}^{'} \\
\end{array}%
\right)
\end{eqnarray}
where the total transfer matrix ${\mathbb M}={\mathbb
M(0,1)}\cdot{\mathbb M(1,2)}$ and ${\mathbb M(j,j+1)}$ are transfer
matrices that couple the wave function in the $j$-th region to the
wave function in the $(j + 1)$-th region. These are given by
\begin{eqnarray}
&&{\mathbb M(0,1)}=\left(%
\begin{array}{cc}
  {\mathbb I}& {\mathbb I} \\
{\mathbb N_{1}^{+}} &{\mathbb N_{1}^{-}} \\
\end{array}%
\right)^{-1}
\left(%
\begin{array}{cc}
  {\mathbb C_{1}^{+}} & {\mathbb C_{1}^{-}} \\
 {\mathbb G_{1}^{+}} & {\mathbb G_{1}^{-}} \\
\end{array}%
\right)\\
&&{\mathbb M(1,2)}=\left(%
\begin{array}{cc}
  {\mathbb C_{2}^{+}} & {\mathbb C_{2}^{-}} \\
  {\mathbb G_{2}^{+}} & {\mathbb G_{2}^{-}} \\
\end{array}%
\right)^{-1}
\left(%
\begin{array}{cc}
  {\mathbb I}& {\mathbb I} \\
{\mathbb N_{2}^{+}} &{\mathbb N_{2}^{-}} \\
\end{array}%
\right)\left(%
\begin{array}{cc}
  {\mathbb K^{+}}&{\mathbb O}  \\
{\mathbb O} &{\mathbb K^{-}} \\
\end{array}%
\right)
\end{eqnarray}
where we have set the quantities
\begin{eqnarray}
&&\left({\mathbb
N_{1}^{\pm}}\right)_{m,l}=\pm\left(\alpha_{m}\right)^{\pm
1}\delta_{ml} \lb{eqn1}
\\
&&\left({\mathbb
C_{\tau}^{\pm}}\right)_{m,l}=\eta_{\tau,l}^{\pm}J_{m-l}\left(\alpha\right)
\\
&&\left({\mathbb G_{\tau}^{\pm}}\right)_{m,l}=\pm
i\Lambda_{l}\xi_{\tau,l}^{\pm}J_{m-l}\left(\alpha\right)
\\
&&\left({\mathbb  K^{\pm}}\right)_{m,l}=\pm e^{\pm
idk_{m}^{'}}\delta_{ml}
\\
&&\left({\mathbb
 N_{2}^{\pm}}\right)_{m,l}=\pm\left(\beta_{m}\right)^{\pm
1}\delta_{ml}\lb{eqn2}
\end{eqnarray}
with the null matrix is denoted by ${\mathbb O}$ and  ${\mathbb I}$ is
the unit matrix. We assume an electron propagating from left to
right with 
energy $\epsilon$ then $\tau=(1,2)$, $\Xi_{0}$ and the null vector $\Xi_{2}^{'}$ read as
\beq
\Xi_{0}=\{\delta_{0l}\}, \qquad \Xi_{2}^{'}=\{b_{m}\}
\eeq
whereas  the vectors of transmitting and reflecting waves
are given by
\beq\lb{TRM}
\Xi_{2}=\{t_{l}\}, \qquad \Xi_{0}^{'}=\{r_{l}\}.
\eeq
From the above considerations,
one can easily obtain the relation
\beq\lb{TRMM}
 \Xi_{2}=\left({ \mathbb M11}\right)^{-1} \cdot\Xi_{0}.
\eeq
The minimum number $N$ of sidebands that needs to be considered is
determined by the strength of the oscillation, $N>\alpha$, and the
infinite series for $T$ can be truncated to consider a finite
number of terms starting from $-N$ up to $N$. Then \eqref{TRMM}
reduces
\begin{eqnarray}
\left(%
\begin{array}{cc}
 t_{-N} \\
. \\
. \\
 t_{-1}\\
 t_{0} \\
t_{1} \\
 .\\
 .\\
 t_{N}\\
\end{array}%
\right)=\left({ \mathbb M11}\right)^{-1}
\left(%
\begin{array}{cc}
 0 \\
0 \\
0 \\
 0\\
 1 \\
0 \\
0 \\
 0\\
 0\\
\end{array}%
\right)
\end{eqnarray}
where  ${ \mathbb (M11)^{-1}}$ becomes now a matrix of order $[2N+1,2N+1]$. This
allows to end up with transmission amplitudes
\begin{equation}\lb{TNK}
t_{-N+k}=M'\left[k+1, N+1\right]
\end{equation}
with  $k=0,1,\cdots, 2N$ and  $M'$ is a matrix element of
${\mathbb( M11)^{-1}}$.
Furthermore,
analytical results are obtained if we consider small values of
{$\alpha={v_1}/ {\varpi}$} and include only the first two sidebands at
energies $\epsilon\pm \varpi$ along with the central band at
energy $\epsilon$.

To explicitly determine
the full expressions of the reflection and transmission coefficients
$R_{l}$ and $T_{l}$,
we use the reflected $J_{\sf {ref}}$ and transmitted $J_{\sf
{trans}}$ {probability} currents
 to write
\begin{equation}\lb{curtr}
  T_{l}=\frac{ |J_{{\sf {tra}},l}|}{| J_{{\sf {inc}},0}|},\qquad R_{l}=\frac{|J_{{\sf {ref}},l}|}{ |J_{\sf {inc,0}}|}.
\end{equation}
Actually, $T_{l}$ is the probability coefficient describing the
scattering of an electron with incident energy $\epsilon$ in the
region 0 into the sideband with quasienergy $\epsilon+l\varpi$ in
the region 2.  Thus, the rank of the transfer matrix ${\mathbb
M}$ increases with the amplitude of the time-oscillating
potential. Now from our Hamiltonian, one can show that
the electrical current density $J$ is given by
\begin{equation}
J= v_{F}\psi^{\dagger}\sigma _{x}\psi
\end{equation}
which is equivalent to write
\begin{eqnarray}
&& J_{{\sf {inc}}, 0}= v_{F} \left(\alpha_{0}+\alpha_{0}^{\ast}\right)
\\
 && J_{{\sf {ref}}, l}= v_{F}r_{l}^{\ast}r_{l} \left(\alpha_{l}+\alpha_{l}^{\ast}\right)
\\
 && J_{{\sf {tra}}, l}= v_{F}t_{l}^{\ast}t_{l} \left(\beta_{l}+\beta_{l}^{\ast}\right).
\end{eqnarray}
These
can be injected in \eqref{curtr} to end up with
the transmission and reflection probabilities
\begin{equation}\lb{tlrl}
  T_{l}= \lambda_{l}\mid t_{l}\mid^{2}, \qquad
  R_{l}=  \kappa_{l}\mid r_{l}\mid^{2}
\end{equation}
where the parameters $\lambda_{l}$ and $\kappa_{l}$ are given by
\begin{eqnarray}
&& \lambda_{l}=
\frac{s_{l}}{s_{0}}\frac{k^{'}_{l}}{k_{0}}\frac{\sqrt{k_{0}^{2}+k^{2}_{y}}}{\sqrt{(k^{'}_{l})^{2}+
  (k_{y}+\frac{d}{l_{B}^{2}})^{2}}}=\frac{\cos\theta_{l}}{\cos\phi_{0}}
\\
 && \kappa_{l}=   \frac{s_{l}}{s_{0}}\frac{k_{l}}{k_{0}}\frac{\sqrt{k_{0}^{2}+k^{2}_{y}}}{\sqrt{k_{l}^{2}+
  k^{2}_{y}}}=\frac{\cos\phi_{l}}{\cos\phi_{0}}.
\end{eqnarray}
 By taking into account of the energy conservation, 
 one can write the two last parameters as
\begin{eqnarray}
&& \lambda_{l}=
\frac{k^{'}_{l}}{k_{0}}\left(1-\frac{s_{l}l\varpi}{\sqrt{(k^{'}_{l})^{2}+
  (k_{y}+\frac{d}{l_{B}^{2}})^{2}}}\right)
\\
 && \kappa_{l}= \frac{k_{l}}{k_{0}}\left(1-\frac{s_{l}l\varpi}{\sqrt{k_{l}^{2}+
  k^{2}_{y}}}\right).
\end{eqnarray}

Due to numerical difficulties,
 we are able to truncate the sums in equations
(\ref{eqn1}-\ref{eqn2}) retaining only the terms corresponding to
the central and first sidebands, namely $l=-1,0,1$. In the forthcoming analysis,
we will analyze each channel separately and sum up their behaviors in the final stage.  Then for
$\alpha=0$ $(v_1=0)$, it remains only  the transmission $t_{0}$ for central
bands that can be analytically determined to obtain
\begin{equation}
t_{0}=\frac{i2\Lambda_{0}\cos\phi_{0}}{e^{idk^{'}_{0}}\chi_{0}}S_{0}
\end{equation}
where different quantities read as
\begin{eqnarray}
&& \chi_{0}=\Gamma_{0}e^{i(\theta_{0}-\phi_{0})}-\Lambda^{2}_{0}\Omega_{0}
-i\Lambda_{0}(I_{0}e^{i\theta_{0}} +G_{0}e^{-i\phi_{0}})
\\
&& S_{0}=\eta^{+}_{2,0}\xi^{-}_{2,0}+\eta^{-}_{2,0}\xi^{+}_{2,0}
\\
&& \Gamma_{0}=\eta^{+}_{1,0}\eta^{-}_{2,0}-\eta^{-}_{1,0}\eta^{+}_{2,0}
\\
&& \Omega_{0}=\xi^{+}_{1,0}\xi^{-}_{2,0}-\xi^{-}_{1,0}\xi^{+}_{2,0}
\\
&&I_{0}=\eta^{+}_{2,0}\xi^{-}_{1,0}+\eta^{-}_{2,0}\xi^{+}_{1,0}
\\
&& G_{0}=\eta^{+}_{1,0}\xi^{-}_{2,0}+\eta^{-}_{1,0}\xi^{+}_{2,0}
\end{eqnarray}
and $T_0$ follows immediately from \eqref{tlrl}.
For $\al\neq 0$, we can proceed as before to derive transmission
amplitudes
\beq
t_{-1}=M'\left[1, 2\right], \qquad t_{0}=M'\left[2, 2\right],\qquad
t_{1}=M'\left[3, 2\right]
\eeq
corresponding to three channels $l=-1,0,1$.
These results will be
analyzed numerically, in terms of different physical parameters,
to underline  
the basic features of our system.

\section{Discussions} 

In this section we present the numerical results for both the
transmission and reflection coefficients, which are shown in
Figures 5, 6, 7, 8, 9, 10, 11, 12, 13 for several parameter values
($\epsilon$, $v$, $\alpha$, $d$). For instance a typical value of
the magnetic field, say $B_{0} = 4 T$, the magnetic length is
$l_{B} = 13 nm$, and $\epsilon l_{B}= 1$ corresponding to the
energy $E = 44 meV$ \cite{martino}. {These typical
values will serve us to normalize the different parameters of our system}.
{Figure 5 illustrates just the transmission of the central band $(l=0)$
as a function of $\alpha$ parameter, which indicates that
transmission guard even allure but with a proportional attenuation
with $\alpha$.}

\begin{figure}[h!]
\centering
\includegraphics[width=10
cm,height=7cm]{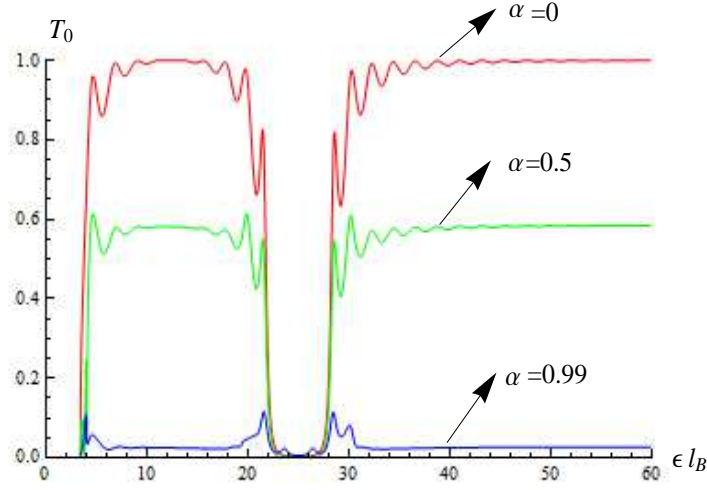}\\
 \caption{\sf{{(Color online)}
  Graphs depicting the transmission $T_{0}$ as a function of energy $\epsilon l_{B}$ for the monolayer graphene barriers
  with $\alpha=\{0, 0.5, 0.99\} $, $d= 1.2l_{B}$, $v l_{B}=25$,  and
 $k_{y}l_{B}=2$.}}
\end{figure}
\begin{figure}[h!]
\centering
\includegraphics[width=10
cm,height=7cm]{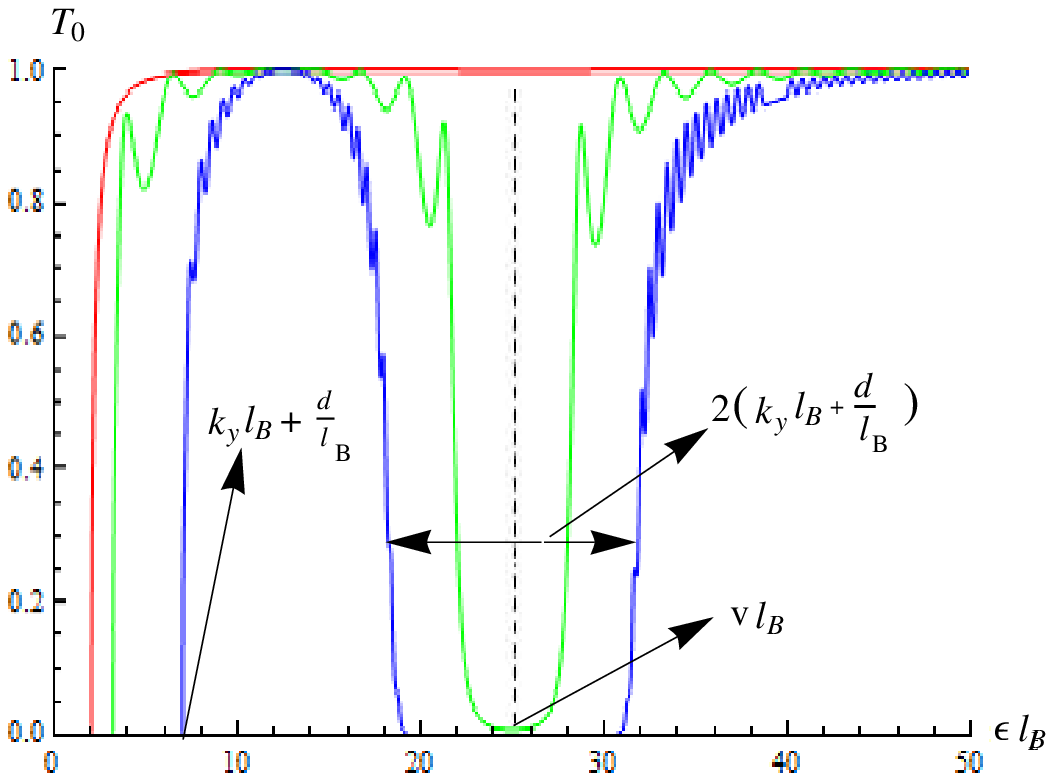}\\
 \caption{\sf{{(Color online)}
Graphs depicting the transmission $T_{0}$ as  function of energy
$\epsilon l_{B}$ for the monolayer graphene barriers with
  $\alpha=0 $, $vl_{B}=25$,
 $k_{y}l_{B}=2$ and $\frac{d}{l_{B}}=0,02$
(red), $\frac{d}{l_{B}}=1.2$ (green) and
$\frac{d}{l_{B}}=5$ (blue).}}
\end{figure}
In Figure 6, the transmission coefficients $T_{l}$ is shown versus the energy
$\epsilon l_{B}$. The quantity $k_{y} l_{B} = m^{\ast}$ plays a
very important role in the transmission of Dirac fermions via the
obstacles created by the series of scattering potentials, because
it is associated with an effective mass of the particle and hence
determines the threshold for the allowed energies. However, the
application of the magnetic field in the intermediate zone where
the barrier oscillates sinusoidally around $v$ with amplitude
$v_{j}$ and frequency $\varpi$ seems to reduce this effective mass
to $(k_{y} l_{B}- l\varpi l_{B})$ in the incidence region while it
increases it to $(k_{y} l_{B} + \frac{d}{l_{B}}-l\varpi l_{B})$ in
the transmission region. The allowed energies are then determined
by the greater effective mass, namely $\epsilon l_{B} \geq k_{y} l_{B} + \frac{d}{l_{B}}-l\varpi l_{B}$.

\begin{figure}[h!]
\centering
\includegraphics[width=10
cm,height=7cm]{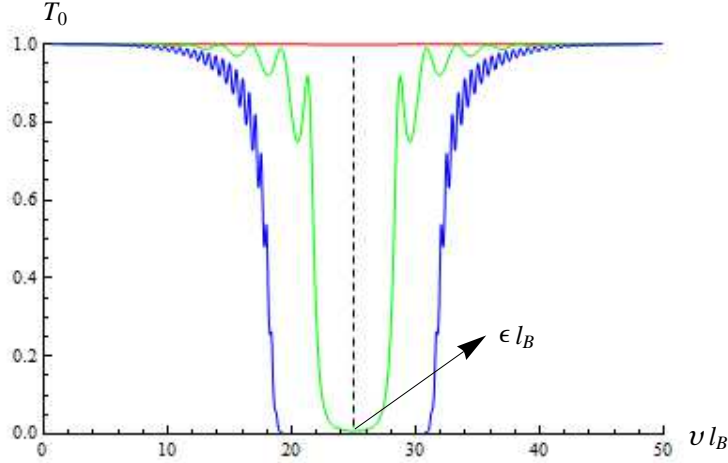}\\
 \caption{\sf{{(Color online)}
  Graphs depicting the transmission $T_{0}$ as a function of potential $v l_{B}$ for the monolayer graphene
  barriers with
  $\alpha=0 $, $\epsilon l_{B}=25$,
 $k_{y}l_{B}=2$ and $\frac{d}{l_{B}}=0,02$
(red), $\frac{d}{l_{B}}=1.2$ (green) and
$\frac{d}{l_{B}}=5$ (blue).}}
\end{figure}
{In Figure 7, one can see that the transmission is depending on   $v
l_{B}$  for  $\alpha = 0$. This shows us how although the
transmission is complete for small widths of the potential and how
 a bowl,  corresponding a total reflection in the vicinity of the
energy of propagation, is wider in terms of the width of the
potential which behaves as the effective mass is added.}

\begin{figure}[h!]
\centering
\includegraphics[width=10
cm,height=6cm]{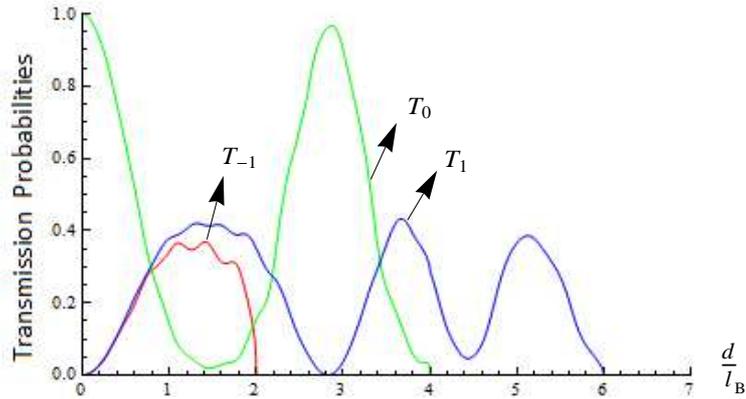}\\
 \caption{\sf{
{(Color online)} Transmission
probability of electrons for central band and first few sidebands
for $\alpha=0.99$ along with that for static barrier as a function
of the barrier width
 $\epsilon l_{B}=5$, $v l_{B}=12$,  $\varpi l_{B}=2$ and
 $k_{y}l_{B}=1$.}}
\end{figure}

\begin{figure}[h!]
\centering
\includegraphics[width=8cm, height=5cm]{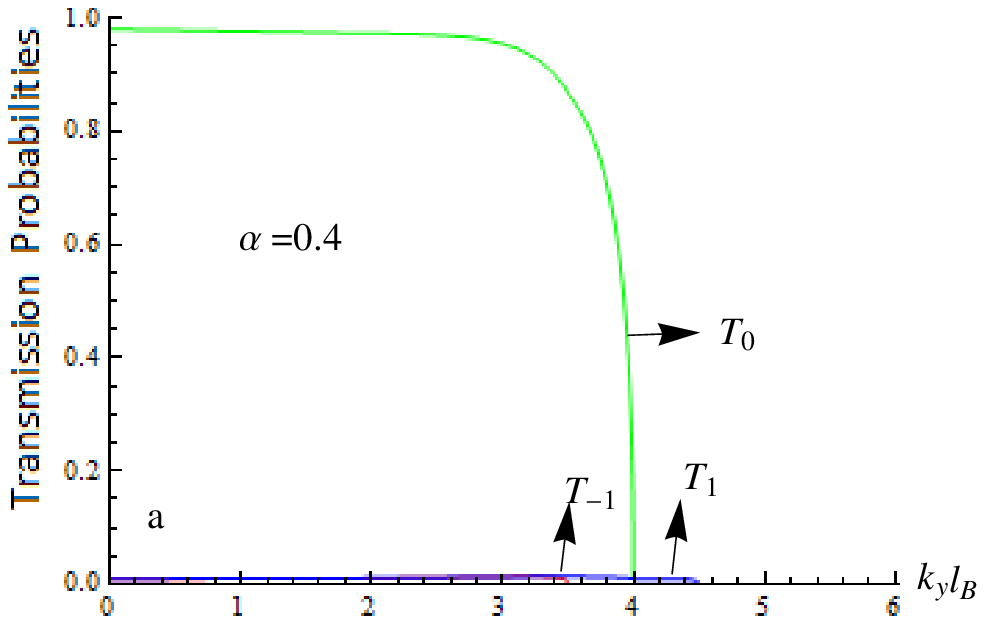}\ \ \ \
\includegraphics[width=8cm, height=5cm]{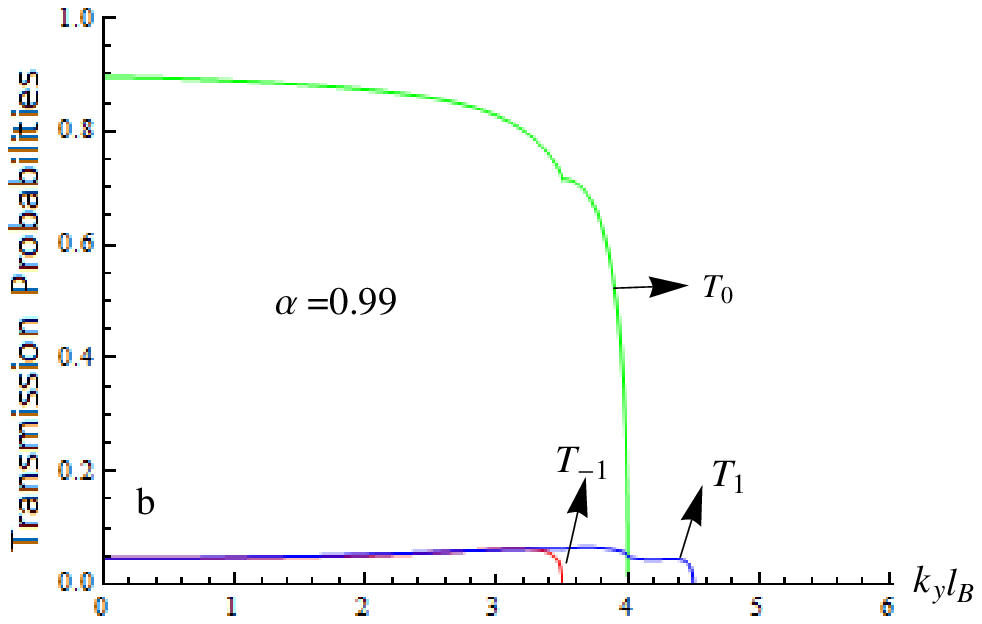}\\
 \caption{\sf{
 {(Color online)} Transmission
probability of electrons for central band and first few sidebands
for a- $\alpha=0.4$ and b- $\alpha=0.99$ along with that for
static barrier as a function of the barrier width $d=1l_{B}$,
$vl_{B}=12$, $\epsilon l_{B}=5$ and
 $\varpi l_{B}=0.5$.}}
\end{figure}

{Both of Figures 8 and 9 show the effects
of $\frac{d}{l_{B}}$ and  $k_y l_{B}$ are similar to the point of
view of the limitations of permitted transmissions. We observe that the two parameters $\frac{d}{l_{B}}$
and $k_yl_{B}$ act as an effective mass respecting, respectively,
the two following relations: $k_{y} l_{B} \leq \epsilon l_{B} -
\frac{d}{l_{B}}+l\varpi l_{B}$ and  $ \frac{d}{l_{B}} \leq
\epsilon l_{B} -k_{y} l_{B}+l\varpi l_{B}$. It should be noted
that the sum of the transmissions of different modes $(l=-1,0,1)$
would never exceeds the unit.}

\begin{figure}[h!]
\centering
\includegraphics[width=8cm, height=5cm]{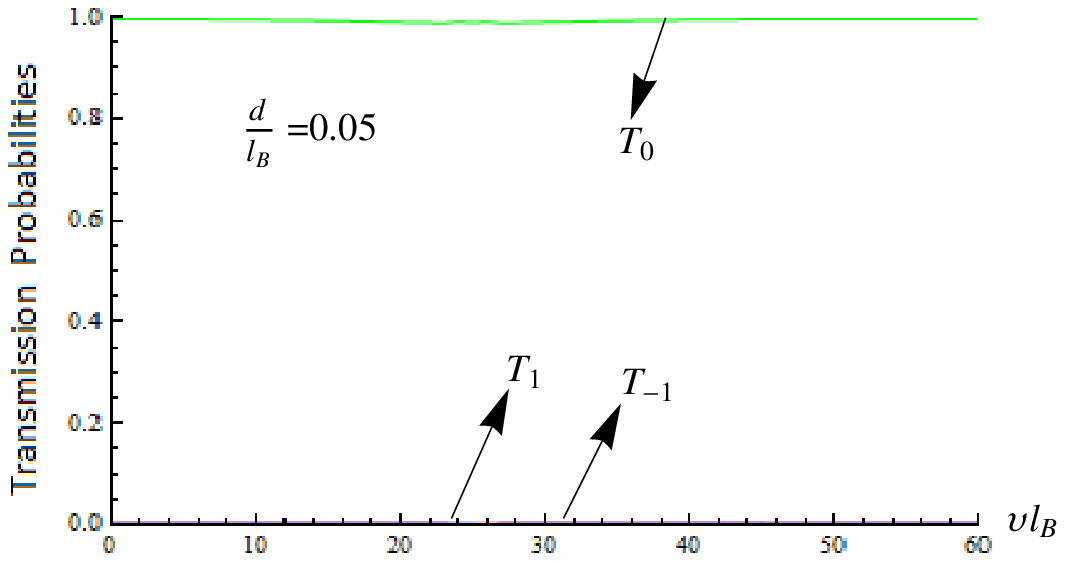}\ \ \ \
\includegraphics[width=8cm, height=5cm]{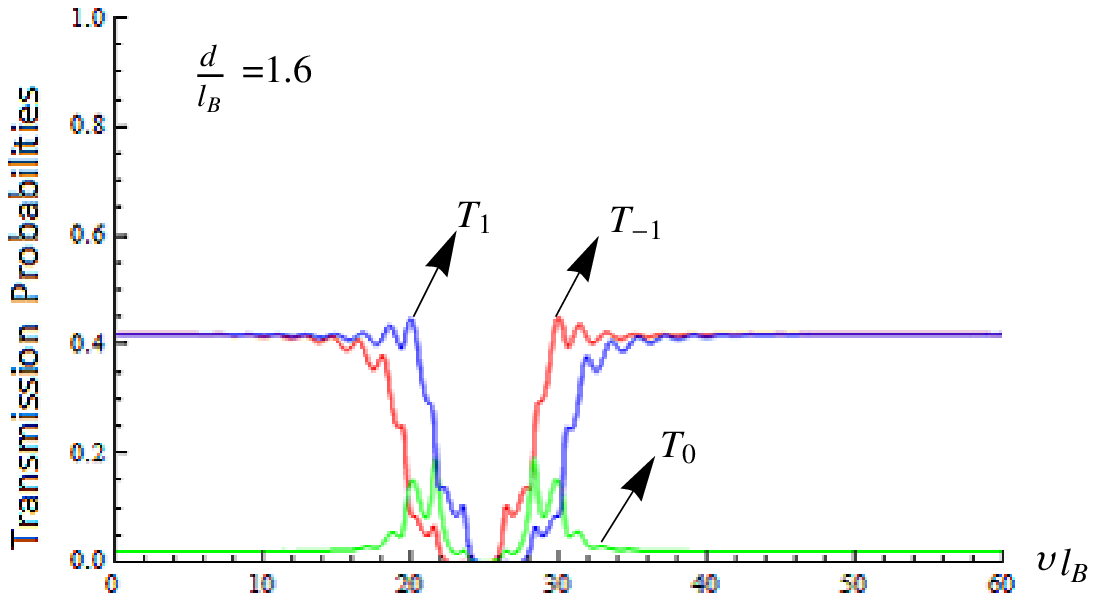}\\
\ \ \ \includegraphics[width=8cm, height=5cm]{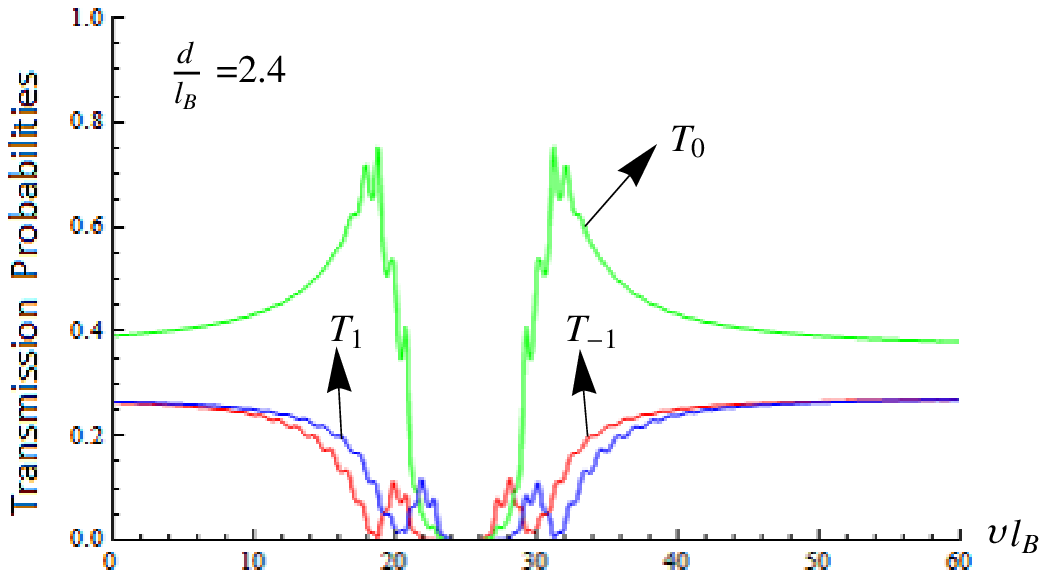}\ \ \ \ \
\includegraphics[width=8cm, height=5cm]{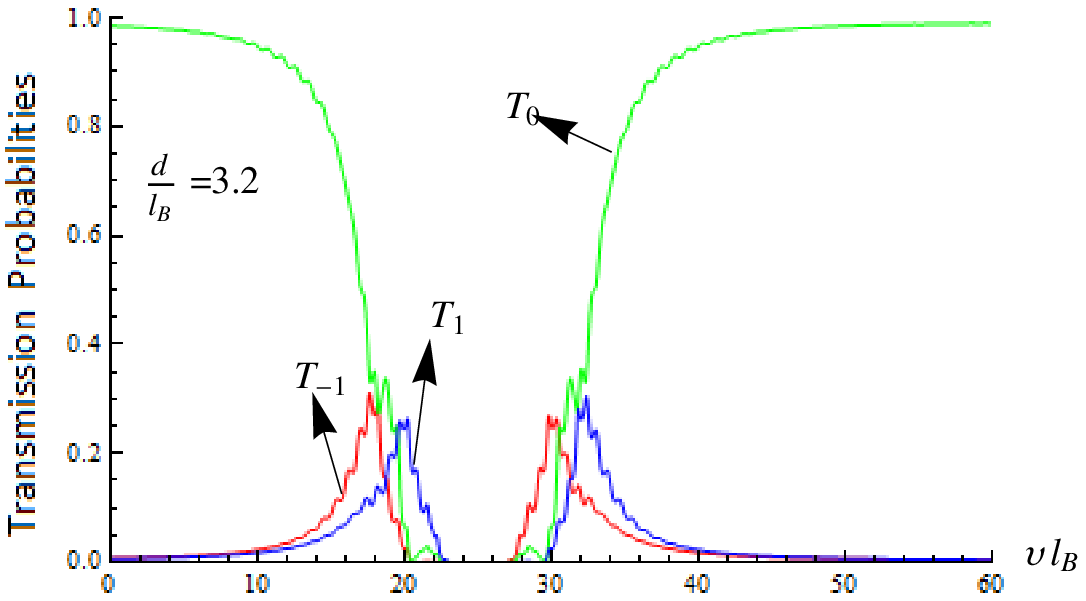}
 \caption{\sf{{(Color online)}
 Graphs depicting the transmission probability for central band $T_{0}$ and first sidebands $T_{\pm1}$ as a function of potential $vl_{B}$ for the
 monolayer
  graphene barriers
  $d=\{0.05l_{B}, 1.6l_{B}, 2.4l_{B}, 3.2l_{B}\}$, with  $\alpha=0.99$, $\epsilon l_{B}=25$, $\varpi l_{B}=2$ and
 $k_{y} l_{B}=2$.}}
\end{figure}

{From Figure 10, one can see that the evolution of the central transmission band and the two lateral
bands is depending on the width from the single oscillating potential
over time accompanied by a magnetic field, and
recognizes four different important phases depending on the
desired applications.
The first phase starts for very small widths which was the
dominance of the central band that is significantly large and that
begins with a total transmission whatever the applied potential.
The second phase comes in second order in which it was the
dominance of the two side bands each of which is symmetrical to
the other relative to an axis of symmetry located at the potential
corresponding to propagation energy. The third phase is similar to
the second but with dominance changing between the central strip
and the lateral strips retaining the sum between the different
transmissions found less than or equal to unity.
In the last phase, the central strip recovers its dominance but
this faith latter with a total reflection from the turns
predicted axis of symmetry and a total distance of the
transmission axis. The transmissions of the two lateral strips are
placed close to the axis of symmetry in which central transmission
is strictly between zero and one such that the sum of all
transmissions does not exceed the transmission unit and each
sideband becomes symmetrical relative to the opposite to the axis
of symmetry.}

\begin{figure}[H]
\centering
\includegraphics[width=8cm, height=5cm]{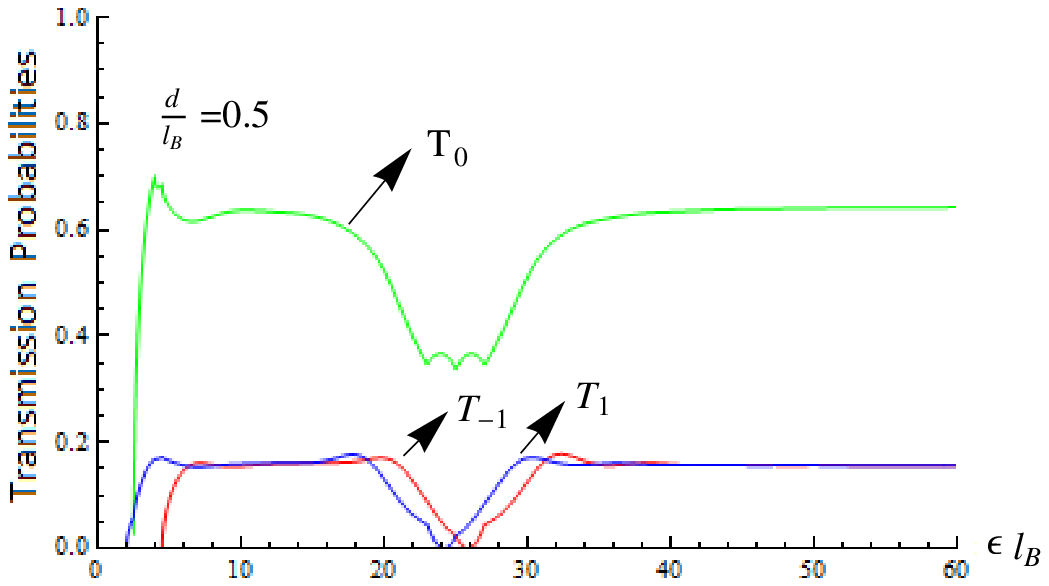}\ \ \ \
\includegraphics[width=8cm, height=5cm]{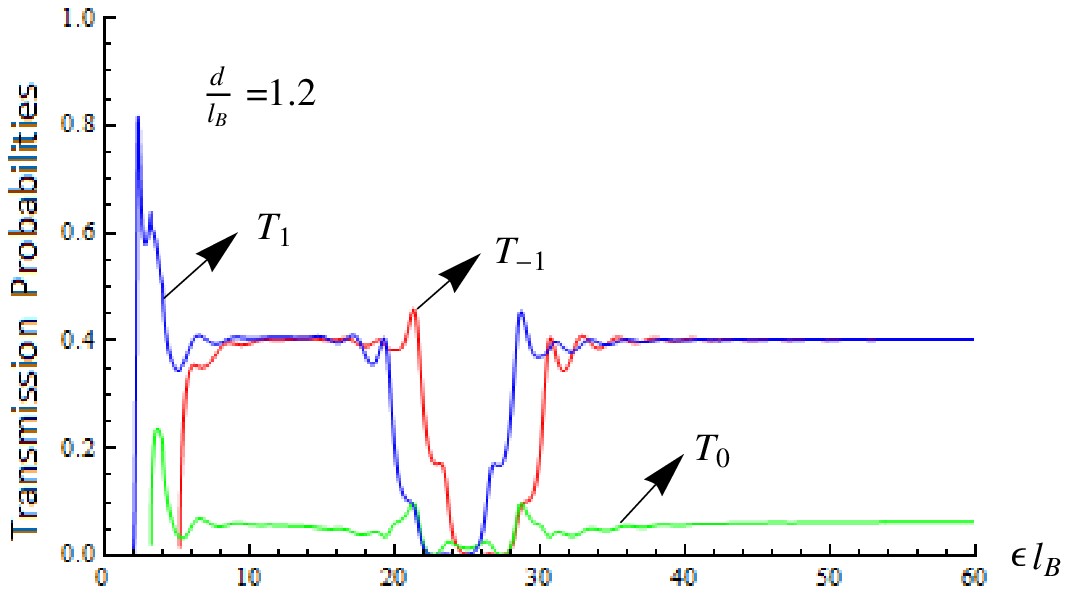}\\
\ \ \ \includegraphics[width=8cm, height=5cm]{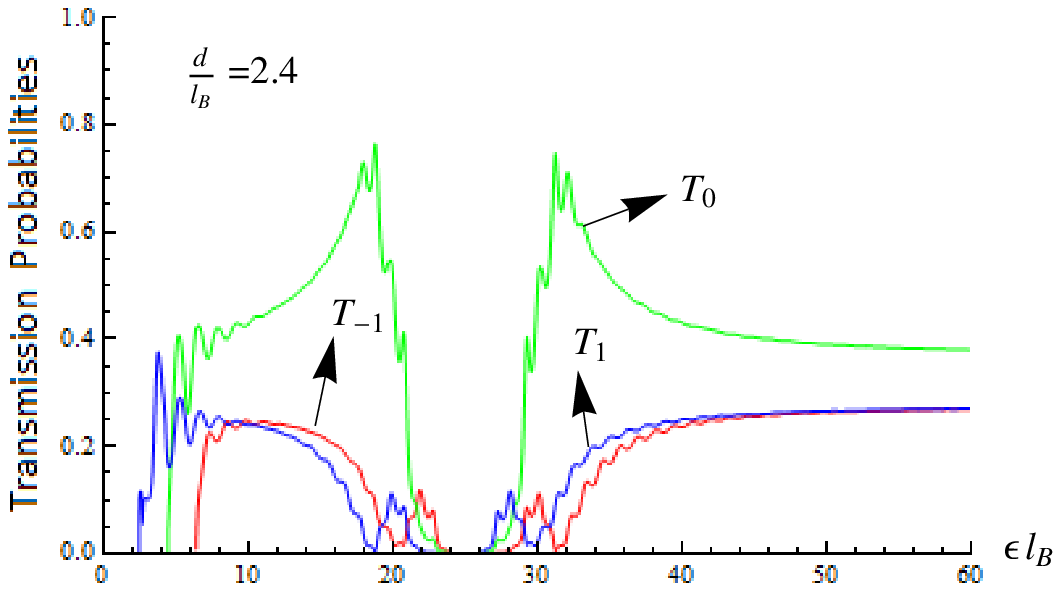}\ \ \ \ \
\includegraphics[width=8cm, height=5cm]{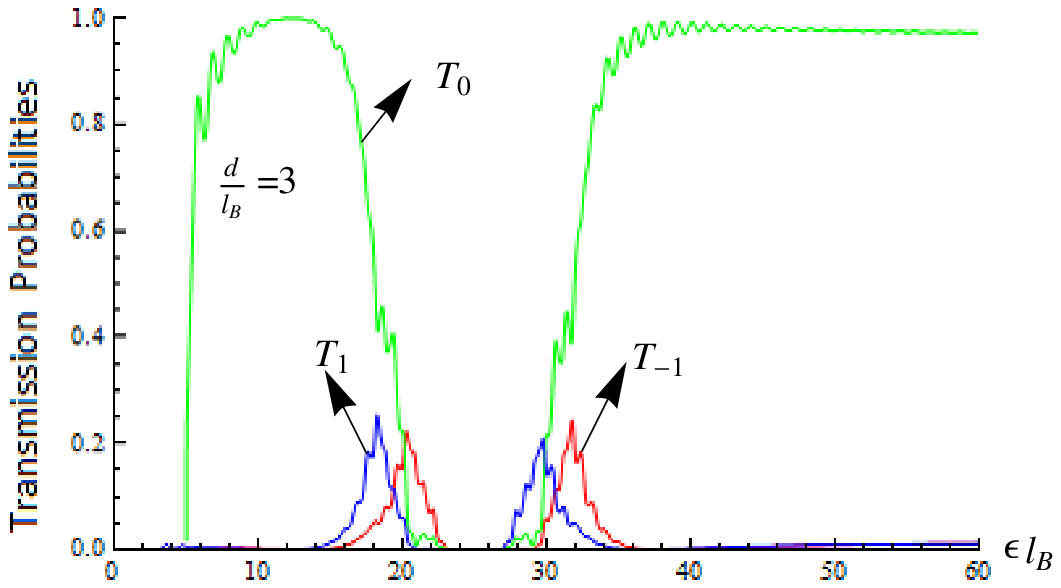}
 \caption{\sf{{(Color online)}
 Graphs depicting the transmission probability for central band $T_{0}$ and first sidebands $T_{\pm1}$ as  function of energy $\epsilon l_{B}$ for the
 monolayer
  graphene barriers
$d=\{ 0.5l_{B}, 1.2l_{B}, 2.4l_{B}, 3l_{B}\}$, with $\alpha=0.99$,
$vl_{B}=25$, $\varpi l_{B}=2$ and
 $k_{y}l_{B}=2$.}}
\end{figure}
{Figure 11 tells us that in the same way the evolution of
the same transmissions, depending on the energy, are as
before. This faith by complying forbidden energies below the
effective mass, namely $\epsilon l_{B} \geq k_{y} l_{B} +
\frac{d}{l_{B}}-l\varpi l_{B}$.}

\section{Total transmission probability}

For static barrier, we know that there is only one transmission probability,
which is function of  
the barrier width. Whereas
%
in the oscillating barrier the total transmission
probability for energy $\epsilon$ is given by the sum over all modes
$l$
\begin{equation}
T=\sum_{l=-\infty}^{l=+\infty}T_{l}.
\end{equation}
In the forthcoming analysis, let us choose the parameters characterizing our system in
such manner that the lateral bands from $l=\pm 2$ and so on  switch off
quickly and the sum of the partial transmissions
\beq
T^{'}=\sum_{-N}^{N}T_l
\eeq
converges significantly to
\beq
T^{'}=T_{-1} + T_0+ T_1
\eeq
which represent from now the total
transmission of all system modes.
With this we will see how the results presented for each mode  previously in different figures
for $T_0$ and $T_{\pm1}$ will be summed up to get the total transmission plots.

\begin{figure}[H]
\centering
\includegraphics[width=8cm, height=5cm]{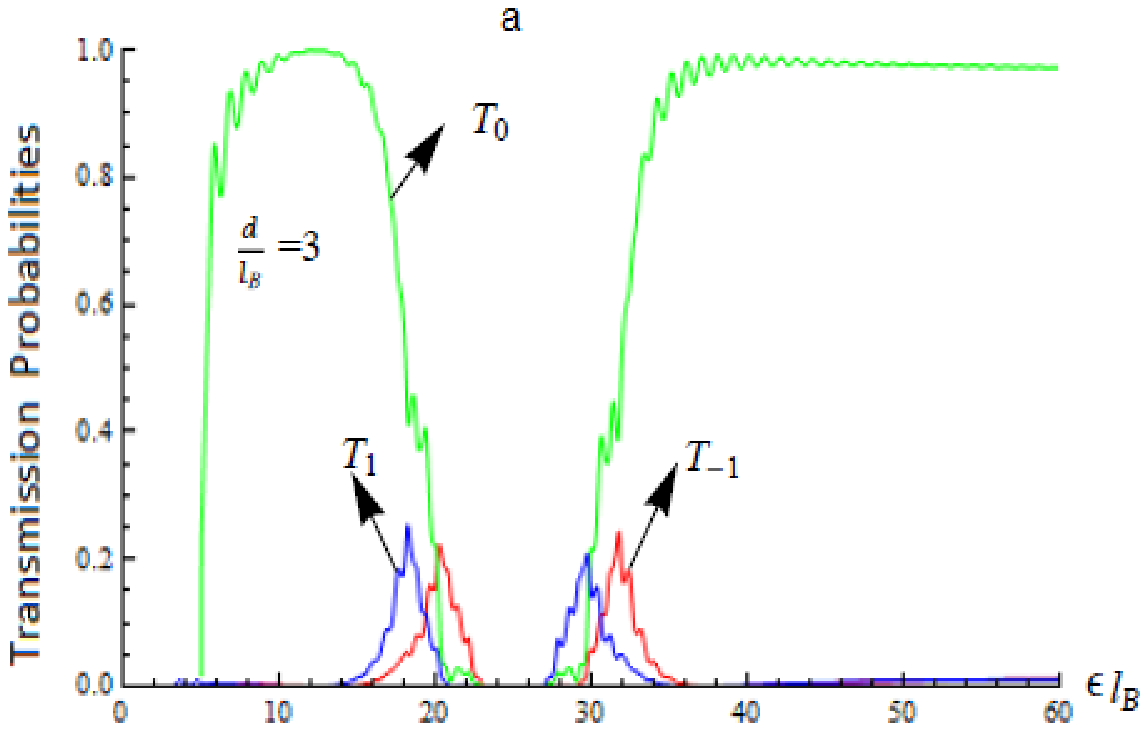}\ \ \ \
\includegraphics[width=8cm, height=5cm]{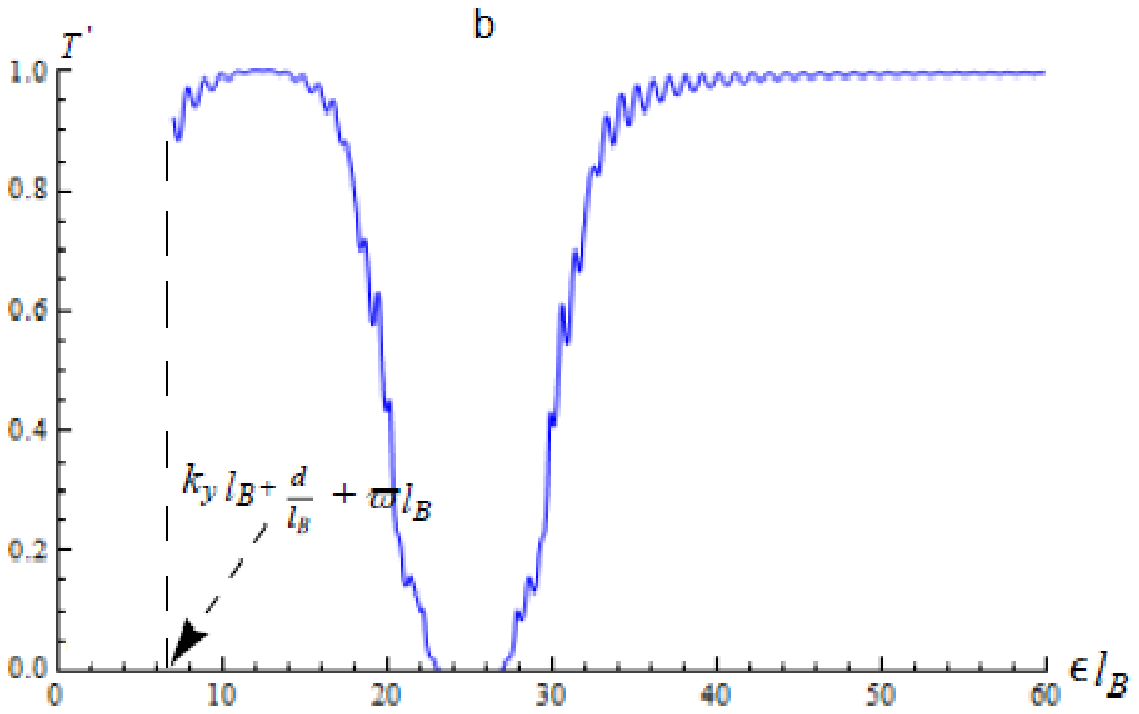}\\
 \caption{\sf{ (Color online)
 a-graphs depicting the transmission probability for central band $T_{0}$ and first sidebands $T_{\pm1}$ as a function of energy $\epsilon
 l_{B}$ and
  b-Graphs depicting the transmission probability for  $T^{'}=T_{-1}+T_{0}+T_{1}$  as a function of energy $\epsilon
  l_{B}$ for the monolayer
  graphene barriers.
  $\frac{d}{l_{B}}=3$, with  $\alpha=0.99$, $v l_{B}=25$, $\varpi l_{B}=2$ and
 $k_{y} l_{B}=2$.}}
\end{figure}

\begin{figure}[h]
\centering
\includegraphics[width=8cm, height=5cm]{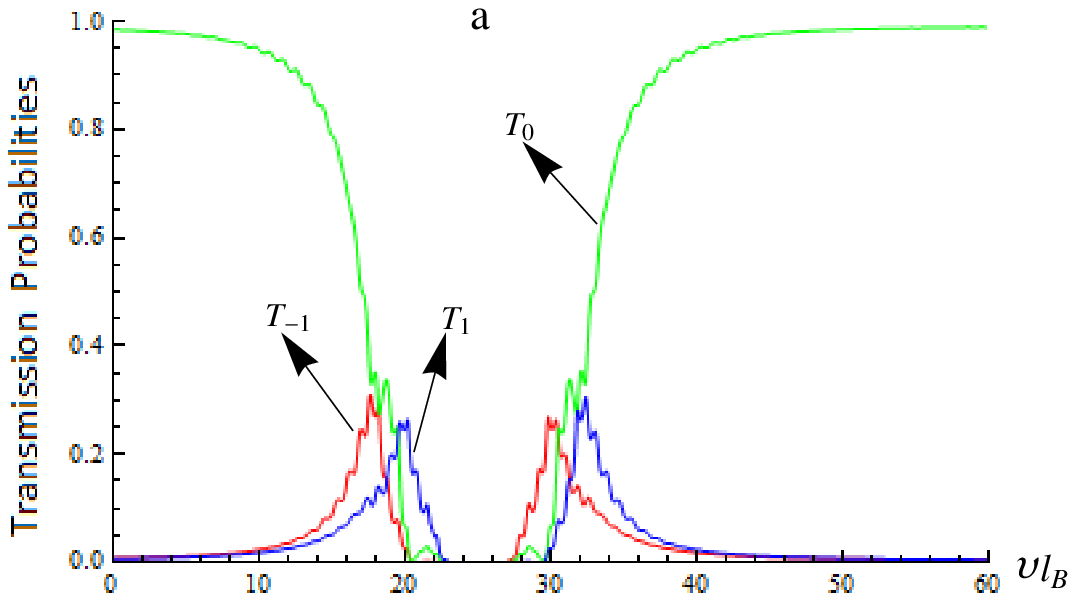}\ \ \ \
\includegraphics[width=8cm, height=5cm]{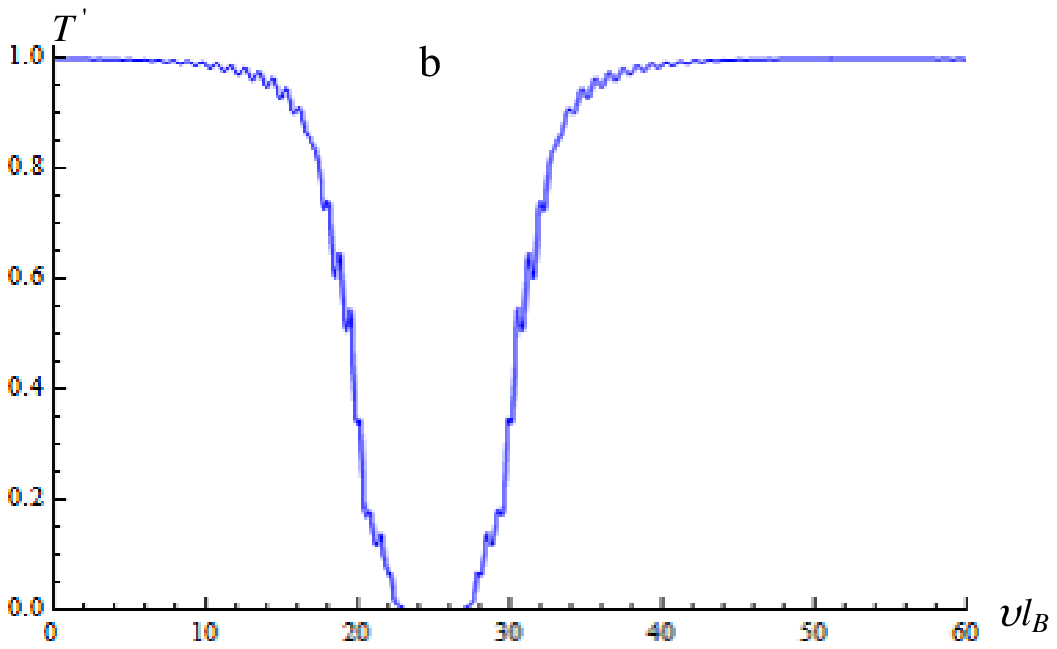}\\
 \caption{\sf{ (Color online)
 a-graphs depicting the transmission probability for central band $T_{0}$ and first sidebands $T_{\pm1}$ as a function of potential $vl_{B}$ and
  b-Graphs depicting the transmission probability for  $T^{'}=T_{-1}+T_{0}+T_{1}$  as a function of potential $vl_{B}$ for the monolayer
  graphene barriers.
  $\frac{d}{l_{B}}=3.2$, with  $\alpha=0.99$, $\epsilon l_{B}=25$, $\varpi l_{B}=2$ and
 $k_{y} l_{B}=2$.}}
\end{figure}

Figure 12 presents $T'$
versus the energy $\epsilon l_{B}$. We notice that the allowed
energies are determined by the greater effective mass, namely
$\epsilon l_{B} \geq k_{y} l_{B} + \frac{d}{l_{B}}+N\varpi l_{B}$.
It is clearly see that the $T_0$ behavior  corresponding central band
is much more dominated than other two remaining bands.

In Figure 13, we present $T'$
in terms of  the potential width  $vl_{B}$. To make a clear comparison with former analysis, we
pick up the last Figure 10 and give the plot
below with the same conditions. This show clearly that we have a fully transmission behavior
that summing up all that obtained for $T_{-1}$, $T_0$ and $T_1$.


\section{Conclusion}

We have considered Dirac Fermions in graphene
subjected to an external magnetic field
and time-dependent potential. The solutions
of the energy spectrum were obtained for three
regions composing the graphene sheet
in terms of different physical parameters
and the Bessel functions. The obtained eigenvalues
are rich so that
we have seen that absorbing energy
quantum $\varpi$ produces interlevel transitions. Because of the Pauli principle an electron with energy
$\epsilon$
can absorb an energy quantum $\varpi$ if only the state with energy $\epsilon+\varpi$ is empty.

 Subsequently, we have studied the effect of both oscillating field
and applied magnetic field on the electron transport through a
single barrier. The time dependent oscillating barrier height
generates additional sidebands at energies $\epsilon  +l \hbar \omega$
$(l=0,\pm1,\cdots)$
in the transmission probability due to photon absorption or
emission. We have observed that perfect transmission probability
at normal incidence (Klein tunneling) persist for harmonically
driven single barrier.

We have investigate how the transmission
probability is affected by various physical parameters, in particular the barrier width, 
energy and oscillation frequency. Thus our numerical results
support the assertion that quantum interference has an important
effect on particle tunneling through a time-dependent
graphene-based single barrier. Since most optical applications in
electronic devices are based on interference phenomena then we
expect that the results  of our computations might be of interest
to designers of graphene-based electronic devices.

\section*{Acknowledgments}

The generous support provided by the Saudi Center for Theoretical Physics (SCTP)
is highly appreciated by all authors. AJ and HB also acknowledge partial support
by King Faisal University and King Fahd University of Petroleum and Minerals, respectively.

\end{document}